\documentclass[conference]{IEEEtran}

\usepackage{indentfirst}
\usepackage{cite}
\usepackage[tbtags]{amsmath}
\usepackage{amsfonts}
\usepackage{amssymb}
\usepackage{dsfont}
\usepackage{times}
\usepackage{subfigure}
\usepackage[english]{babel}
\usepackage{bbm}
\usepackage{dsfont}
\usepackage{graphicx}
\usepackage{array}
\usepackage[normalem]{ulem}
\usepackage{epsf}
\usepackage{psfrag}
\usepackage{comment}
\usepackage{color}
\usepackage{xcolor}
\usepackage[all]{xy}
\usepackage{booktabs}
\usepackage{algorithmic}
\usepackage[ruled,vlined]{algorithm2e}
\usepackage{caption}

\SetKwFunction{Fn}{\textbf{Function}}%
\SetKwFunction{Dd}{D2Didle}%
\SetKwFunction{Nb}{NoBackoff}%
\SetKwFunction{Bk}{BackOff}%

\newcommand{\figww}{0.99\columnwidth}

\begin{document}

\title{Content-based Cognitive Interference Control for City Monitoring Applications in the Urban IoT}
\author{\large Sabur Baidya and Marco Levorato\\
\normalsize The Donald Bren School of Information and Computer Science, UC Irvine, CA, US\\
\normalsize e-mail: \{sbaidya,~levorato\}@uci.edu\vspace{-0.5cm}}
\date{}

\maketitle

\pagestyle{empty}
\thispagestyle{empty}

\pagestyle{empty}
\begin{abstract}
In the Urban Internet of Things (IoT), devices and systems are interconnected at the city scale to provide innovative services to the citizens. However, the traffic generated by the sensing and processing systems may overload local access networks. A coexistence problem arises where concurrent applications mutually interfere and compete for available resources. This effect is further aggravated by the multiple scales involved and heterogeneity of the networks supporting the urban IoT. One of the main contributions of this paper is the introduction of the notion of \emph{content-oriented} cognitive interference control in heterogeneous local access  networks supporting computing and data processing in the urban IoT. A network scenario where multiple communication technologies, such as Device-to-Device and Long Term Evolution (LTE), is considered. The focus of the present paper is on city monitoring applications, where a video data stream generated by a camera system is remotely processed to detect objects. The cognitive network paradigm is extended to dynamically shape the interference pattern generated by concurrent data streams and induce a packet loss trajectory compatible with video processing algorithms. Numerical results show that the proposed cognitive transmission strategy enables a significant throughput increase of interfering applications for a target accuracy of the monitoring application.
\end{abstract}

\section{Introduction}

The Internet of Things (IoT) is a recent communication paradigm that enables the interconnection and interoperation of everyday life objects, equipped with transceivers for digital communication and suitable protocol stacks~\cite{atzori2010internet}. 
An emerging trend proposes the integration of the IoT paradigm to the urban environment~\cite{IoT,bellavista2013convergence}. Instrumental to the realization of the urban IoT is the real-time exploitation of a new range of data that can be collected by distributed sensors and then uploaded to computation services, which may be located at the edge of the network~\cite{edge_mining_2013}. The urban IoT can interconnect a plethora of public systems and services such as surveillance and monitoring, transportation and parking, and health care systems. The interconnection of these systems to central or local data processing and control centers can indeed improve their efficiency, and can stimulate the creation of new services to the citizens.

However, the capillary interconnection of a large number of devices and systems at a city-wide scale poses several technological challenges to overcome. First, these architectures will use local access networks (\emph{e.g.}, cellular networks, and private and public wi-fi) to enable interoperation. A coexistence problem, then, arises with traditional and new applications. Additionally, the algorithms processing information may impose stringent short-term QoS requirements to the network, which are difficult to meet in the complex, dynamic and heterogeneous environment characterizing the urban IoT.


We contend that the notion of cognitive network~\cite{mitola} can play an important role in the design of networking technologies capable of analyzing the surrounding network environment and dynamically adapt channel access and transmission protocols to facilitate the coexistence of concurrent applications in the urban IoT. In this paper, we propose to use the intelligence implemented in the cognitive terminals to create \emph{content-oriented} interference control to concurrent data streams processed in real-time by urban IoT systems.

We specifically focus on city monitoring applications, where video data streams from surveillance camera systems are processed by real-time edge computational resources to perform object detection ~\cite{hengstler2007mesheye}. This choice is motivated by the relevance of these systems as a component of smart transportation (\emph{e.g.}, traffic monitoring) and surveillance systems~\cite{bramberger2004real} in the urban IoT, as well as by the inherent challenges associated with transmitting video streams over wireless~\cite{fitzek2001mpeg}.
We consider a heterogeneous network scenario, where Device-to-Device (D2D) communications coexist with Long-Term Evolution (LTE) cellular communications in the same bandwidth~\cite{5350367}. We contend that this scenario is particularly suited to urban IoT systems, where local and city-wide data exchange are needed.

By means of detailed network and processing simulations, we demonstrate that the interference from other concurrent applications may severely impair the performance of algorithms processing the video streaming in real time. We propose to use cognition to alleviate this issue. The cognitive terminals - D2D transmitters in the considered scenario - shape their transmission process to generate an interference pattern compatible with the city monitoring application, while maximizing their own achieved throughput. The cognitive terminals exploit the structure of the temporal encoding techniques implemented at the sources of the video streams to minimize the impact of packet loss on video reconstruction. In particular, the interference generated by the cognitive terminals is reduced when a sequence of LTE packets associated with a reference frame is being transmitted. In fact, the effect of packet loss incurred in such frames propagates to the differentially encoded frames in the same group, that is, interference in that period impacts a larger portion of the transmission. Increasingly, the correlated damage in consecutive frames, such that generated by errors in reference frames, severely impacts video processing. 

The implementation of the proposed scheme requires the extension of the LTE protocol stack of current NS-3 simulator to mimic the scenario of D2D communication as an LTE underlay, coexisting with end-to-end video streaming over LTE networks and implement the content-based  Medium Access Control techniques. Numerical results from the simulations show that, for a desirable object detection probability, the proposed transmission strategy grants a significant throughput increase to applications coexisting with the video stream. The harvested channel capacity can be used to support local exchange of information or data upload to the cloud, thus facilitating computation in the urban IoT.

The rest of the paper is organized as follows. Section~\ref{sec:net} describes the network scenario considered in this paper. In Section~\ref{sec:app}, we describe the city monitoring and surveillance application. Section~\ref{sec:cogint} presents the proposed cognitive interference control strategies. In Section~\ref{sec:numres}, we show experimental simulations and evaluations. Section~\ref{sec:ccl} concludes the paper.

\section{Network Scenario}
\label{sec:net}

4G LTE technology is a key component of the overall network infrastructure supporting the urban IoT. However, the proliferation of local services (\emph{e.g.}, social networking) in recent years led to the inclusion of D2D communications in recent standards of 3GPP~\cite{3gppps} as Proximity Services (ProSe). Thus, in context of LTE network, D2D devices instead of transmitting via eNodeB, perform short-range data exchanges with other devices using the LTE frequency band. In order to have low infrastructure overhead and complexity, D2D is used as underlay to LTE communications~\cite{5350367,yu2009performance}, that is, D2D and LTE shares part of same frequency spectrum used by ongoing LTE communications. Currently, two modes of D2D communication are proposed for the standard; Network independent direct ProSe where User Equipments (UEs) in proximity can communicate with each other even if they are not served by E-UTRAN ((Evolved Universal Terrestrial Radio Access Network); the other one is network-assisted D2D communication where both of the UEs are supported by E-UTRAN. 

\begin{figure}[t!]
\centering
\begin{minipage}[b]{0.50\textwidth}
	\hspace{-5mm}
	\includegraphics[width=\textwidth]{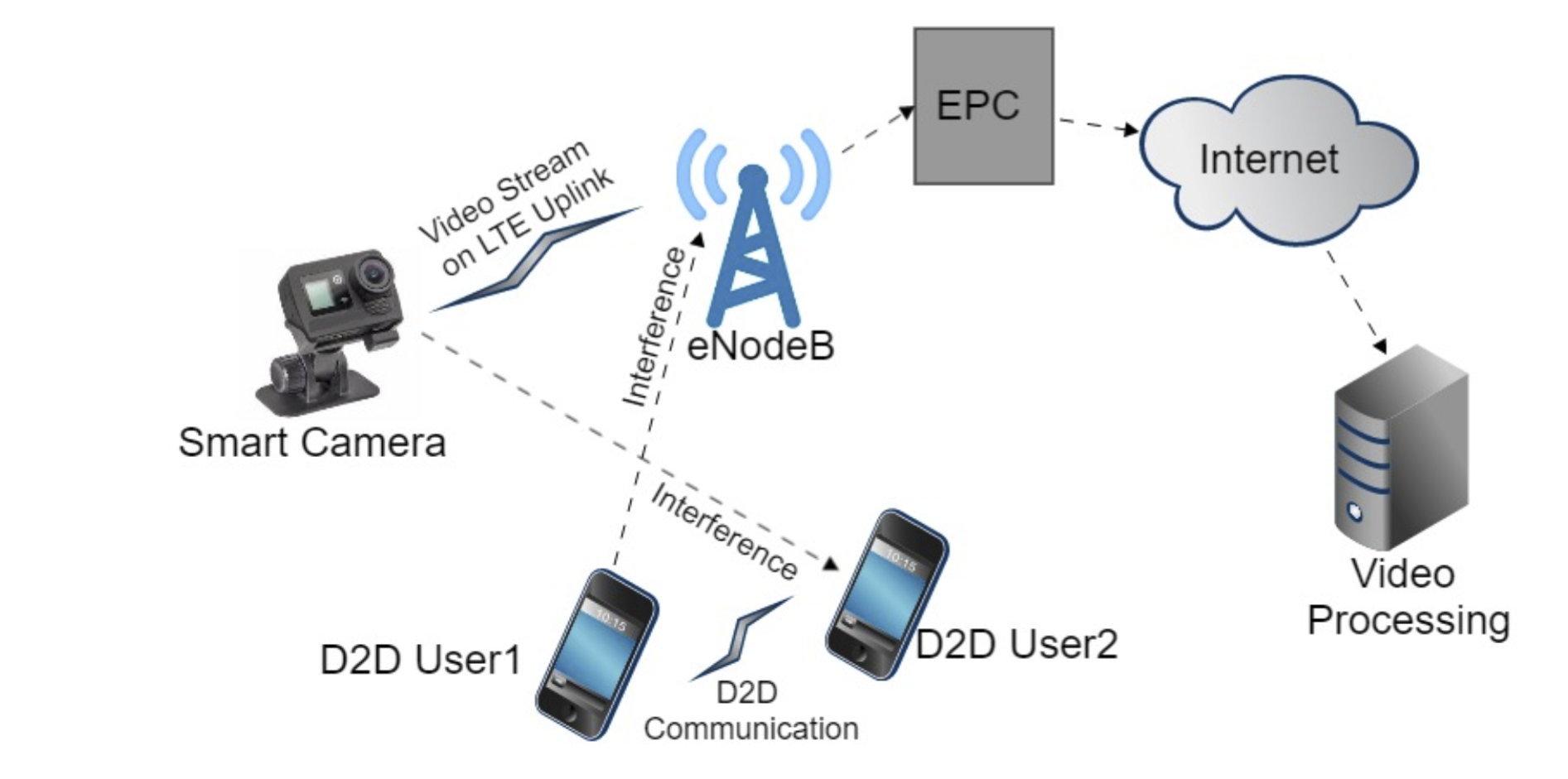}
	\captionsetup{width=1.0 \textwidth}
	\caption{D2D underlay on LTE network with Video \\ Streaming Application on Uplink}
	\label{fig:in1}
	\vspace{-1.5em}
\end{minipage}
\end{figure}

Although D2D communications can be realized using other wireless technologies, such as WiFi and Bluetooth, the multi scale networking architecture created by the composition of D2D and LTE communications perfectly matches the data traffic and bandwidth requirements of urban IoT architecture, where local and global data exchange is needed to support computing and system interconnection. This is especially essential in the light of emerging edge and fog computing paradigms. However, controlling mutual interference in such heterogeneous network scenarios is extremely challenging. Prior works proposed scheduling and interference control strategies that aim at the limitation of the Signal to Interference plus Noise Ratio (SINR) at the LTE receiver~\cite{phunchongharn2013resource,yu2009performance,7127550}.

However, these techniques require instantaneous channel knowledge, and they may not harvest all the available bandwidth in this complex coexistence scenario. Our objective is to shape the transmission process of the D2D link to maximize its throughput while generating an interference pattern compatible with the specific coexisting application (data stream and processing algorithm) supported by the LTE network. Importantly, Our approach only requires long-term statistical knowledge of the wireless channel to efficiently \emph{shape} the transmission process. Also, it takes the content into account for interference control and it only needs LTE control messages that indicates type of data as a preamble before specific type of data packet transmission.

\subsection{Uplink Transmission and Interference scenario}

We consider a scenario where FDD-LTE uplink spectrum is shared with D2D communication.  Conforming with the 3GPP standard for proximity services~\cite{3gppps}, we choose a topology where an end host is transmitting real-time data on the uplink of LTE to the Internet for computation and processing, and two mobile devices in proximity are connected with each other with  network assistance for D2D communication as shown in the Fig.~\ref{fig:in1}. Here, we consider network-assisted D2D~\cite{3gpppr, gunes2014}, where the LTE operator assists in establishing the D2D connection through authentication and authorization over E-UTRAN. Once the connection is established, the data communication happens via direct radio rather than using the E-UTRAN infrastructure which can be used for sending occasional control and signaling messages. 

In the proposed scheme, the UE is streaming real-time video over LTE network on uplink which is transmitted via Serving Gateway (SGW) and  Packet Data Network Gateway (PGW) in the Evolved Packet Core (EPC) network towards remote host. Here we are primarily interested in uplink transmission by E-UTRAN. When the RRC (Radio Resource Control) layer of UE wants to transmit a Packet Data Unit (PDU) on uplink, it sends to underneath PDCP (Packet Data Convergence Protocol) layer which in turn sends to RLC (Radio Link Control) sublayer. The RLC reports its buffer status to the underlying MAC which then sends control message to the eNodeB MAC scheduler. The scheduler allocates Resource Blocks (RBs) for data transmission based on the received signal strength which varies with channel quality, noise and interference. As per the standard, for LTE uplink, the SINR can be measured by control signal e.g. SRS (Sounding Reference Signal) or data signal e.g. PUSCH (Physical Uplink Shared Channel) at eNodeB MAC layer which then calculates a Channel Quality Index (CQI) and reports to the scheduler. The scheduler chooses a suitable Modulation and Coding Scheme (MCS) and notifies to the UE MAC through downlink control information. The UE MAC then transmits the PDU(s) to send via physical layer based on the data rate supported by the MCS assigned.

When the D2D communication interferes with the LTE uplink, the channel quality degrades and the LTE receiver will more likely fail to decode the PDU. Based on the interference level, appropriate CQI and MCS values are chosen as per the standard~\cite{3gpp1}. If the interference is high, the scheduler chooses the lowest MCS which may not be sufficient for transmission of the PDUs and hence causes packet loss.

\section{Application Scenario: Surveillance Camera and City Monitoring}
\label{sec:app}

In context of urban IoT, city monitoring and public safety applications become integral part of services enabled by the interconnection of multiple systems. These systems are inherently distributed systems, where local and edge processing are integrated to increase performance while reducing network traffic load. Herein, we consider a scenario where video stream from a surveillance camera is processed~\cite{chellappa2008} by edge-computing and sent to data management system. This can be extended in future to smart camera systems, where the cameras locally coordinate to determine encoding and transmission parameters of the data stream uploaded to the processing centers~\cite{calavia2012}.

\subsection{Video Streaming}

Video is a sequence of images produced at a given rate. Unlike images, which only have a spatial component, a video has a temporal component as well. Therefore, compression is performed both along the spatial and temporal dimensions. 

\noindent
\underline{Spatial compression}:
The individual pictures are broken into ``macroblocks'' which are transformed by Discrete Cosine Transform (DCT) from space domain to frequency domain. The basic MPEG-4 use $8{\times}8$ DCT where as H.264/MPEG-4 AVC(Advanced Video Coding) standard uses a $4{\times}4$ DCT-like integer transform. The transformed data, then, is quantized and encoded by entropy coding. 

\noindent
\underline{Temporal compression}:
Temporal compression exploits the significant similarities that may interest pictures within the video stream captured at close time instant. This gives the opportunity to encode the video in less number of key (reference) frames and more number of compressed predicted frames following the reference frame. The decoder uses the preceding reference frame information to decode each of the predicted frames. 

When an encoded frame is damaged, due to spatial compression, it affects the transform coefficients which leads to the multifold corruption in the decoded image content and resolution. The spatial propagation of errors may create artifacts that are detected as objects, or impair the ability of the algorithm to detect existing objects. On the other hand, due to temporal compression, the effect of the corruption varies depending on whether the damaged frame is a reference frame or not. If a reference frame is corrupted, the effect propagates through the entire Group of Pictures (GoP). Alternately, if a predicted frame is damaged, the effect is not so severe compare to loosing a reference frame, as it may loose some information regarding motion vectors but key features in following frames are reconstructable. Here we consider H.264/AVC encoding which uses a Group of Pictures (GoP) for predictive coding of frames. The GoP starts with intra-coded frame (I-frame) followed by a number of inter coded frames \emph{e.g.}, Predicted frames (P-frames), Bi-directional predicted frames (B-frames) and can be of constant size or variable size. 

The video streaming is done by packetizing the video data in a transport stream packet of fixed size 188 bytes. This data is then encapsulated by lower level protocols and sent over the network. The most popular video streaming protocols are HTTP Live Streaming (HLS) and Dynamic Adaptive Streaming over HTTP (DASH). If a packet is lost, the underlying TCP retransmits the packet. However, this setting may result in excessive delay in wireless environments, where channel variations and uncertainty induce non-negligible packet loss. On the other hand, protocols not providing packet retransmission such as Real time protocol (RTP) and UDP, packet loss are not recovered. In our proposed scheme, we reduce the effect of packet loss by cognitive interference control without using any retransmission overhead.

\subsection{Video Processing}

We use Object detection as a metric for measuring the quality of the video. Object detection has been one of the primary area of research in computer vision. Most of the object detection algorithms extract features from the image frames and employ a matching or detection algorithm with respect to reference image reported. Popular object detection algorithms~\cite{viola2001,dalal2005, svm2011} include feature-based object detection, Viola-Jones object detection, object detection based on SVM (Support Vector Machines) and HOG (Histogram Oriented Gradient) features, Image segmentation and blob analysis. In this paper, for illustrative purposes we used the Robust Features (SURF)~\cite{baya2008} based object recognition technique to recognize the point objects in each frame of a video with respect to a reference image. So, while interference corrupts the video and affects the object detection, we measure the object detection probability of the video as: 
\begin{equation}
P_{det} = \frac{F_{rcv}}{F_{ref}}
\end{equation}
where $F_{rcv}$ and $F_{ref}$ are number of correctly detected surface objects in the received and reference video respectively such that $\{F_{rcv}, F_{ref}\} \in F $ where F is the Feature set detected in the original uncorrupted video.

\section{Cognitive Interference Strategy}
\label{sec:cogint}

We take the approach of controlling the D2D transmission probability with respect to a given D2D transmission power in order to control the interference on the video transmission over LTE, achieving tolerable packet loss. We focus on topologies where the transmission power of the LTE user (transmitting the video stream) is sufficient to guarantee low packet failure probability in the absence of interference. 

\begin{figure}[t!]
	
	\begin{minipage}[b]{0.65\textwidth}
	\raggedright
	\hspace{-6mm}
	\includegraphics[width=90mm]{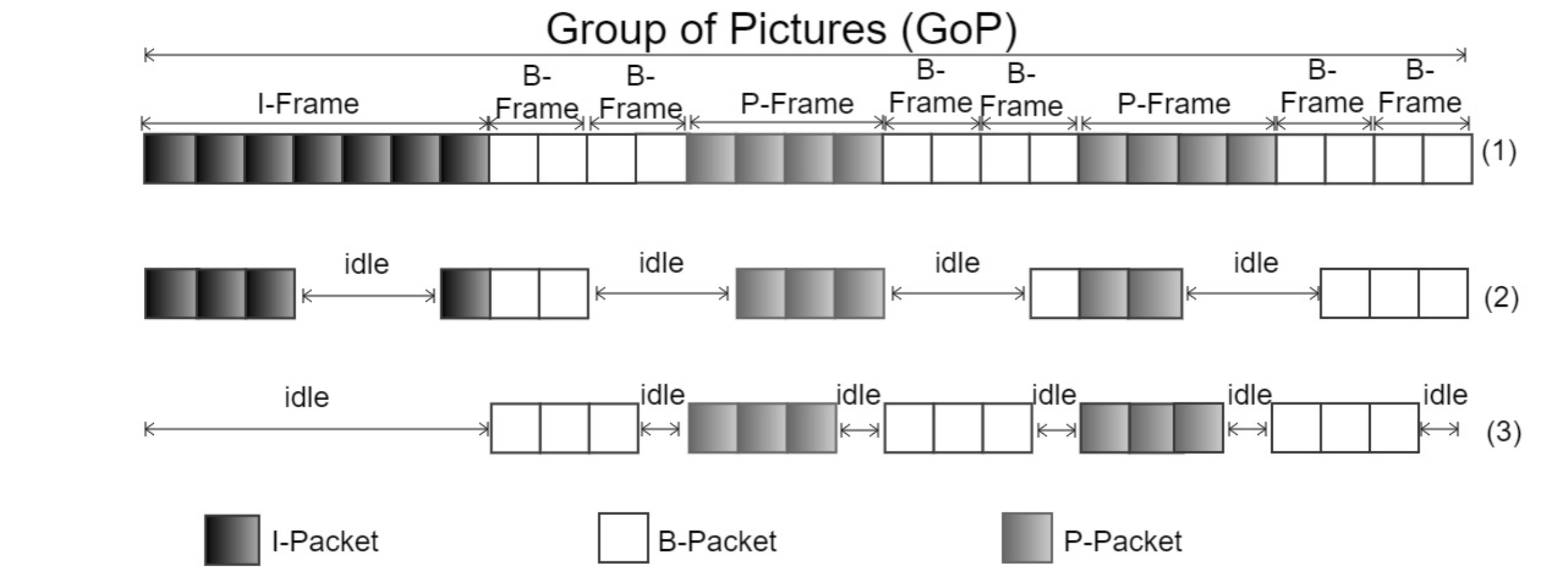}
	\captionsetup{width=1.0\textwidth}
	\raggedright
	\hspace{-6mm}
	\caption{Cognitive Interference Strategies: (1) Video \\packet flow (2) Fixed D2D Tx probability (3) FDTP}	
	\label{fig:prtcl}	
	\end{minipage}
\end{figure}

We propose a novel cognitive transmission technique, where the transmission power and channel access scheme of the D2D link are based on the structure of the video data stream. In particular, we define a strategy that differentiates between groups of packets in the LTE video stream to improve the D2D throughput while limiting its impact on the ability of the monitoring and object detection algorithms. Our transmission strategy is based on the observation that damage to groups of LTE packets transporting a reference frame impacts the entire GoP. Transmission activity by the D2D link within a temporal period where packets associated with reference frames are transmitted, generates a larger \emph{effective interference} to the video quality with respect to the same activity performed during differential frames transmission. Increasingly, the effective interference generated by damage of reference frame is further enhanced by the error patterns induced to the video stream, which may result in the detection of non-existing objects.

The proposed technique only requires a statistical knowledge of the network channel coefficients, as opposed to the fine-granularity channel information required to control the SINR slot by slot. This a priori information is used by the D2D terminal to compute its transmission power, which determines the success probability at the D2D and LTE receivers conditioned on transmission. Then, the D2D terminal \emph{shapes} its access strategy based on the video frame being transmitted by the LTE user. We call this as Frame Dependent Transmission Probability (FDTP).

 \begin{figure}[ht!]
\centering
\begin{minipage}[b]{0.35\textwidth}
	\vspace{-2mm}
	\hspace{-0mm}
	\includegraphics[width=\textwidth]{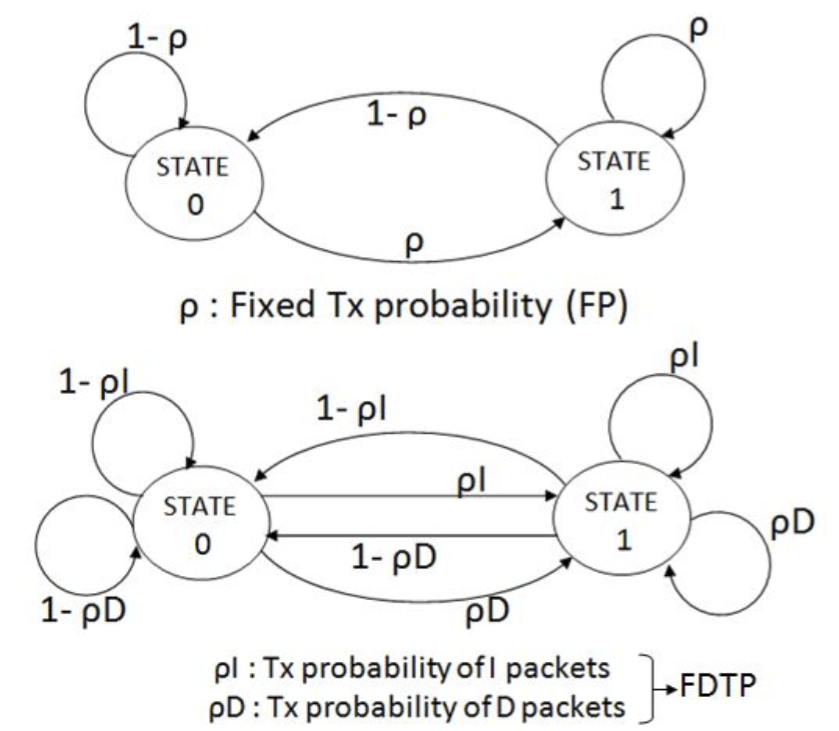}
	\captionsetup{width=1.2 \textwidth}
	\caption{Finite State Machines for the interference control streategies}
	\label{fig:intfsm}
	\vspace{-1.5em}
\end{minipage}
\end{figure}

We compare FDTP scheme against a simple scheme where the D2D node determines transmission power and access probability $\rho$ to induce a fixed average failure probability to LTE packets. Then, in each slot the D2D terminal transmits with probability $\rho$ with a predetermined transmission power. We refer to this strategy as Fixed Probability (FP). Fig.~\ref{fig:prtcl} shows a graphical representation and Fig.~\ref{fig:intfsm} shows the finite state machines for the proposed schemes. In the following, we describe the proposed FDTP strategy.

\vspace{2mm}
\noindent
\underline{\emph{Frame Dependent Transmission Probability (FDTP)}}:

\noindent
The D2D transmitter switches transmission access modes (LOW and HIGH) depending on whether the LTE is transmitting reference (I) or differential (B/P) frames. These modes correspond to transmission probabilities equal to $\rho_I$ and $\rho_D$, respectively. 

We implement this strategy in the LTE network scenario we described earlier. The LTE UE transmitting the video stream, sends a preamble message with the control information on the uplink to the eNodeB about the video frame type (I, B/P) before sending new type of frame(s). As a GoP contains a series of frames and each frame contains hundreds of packets, the control message overhead for the preamble is much less compared to total number of packets. As we consider that the D2D devices are network assisted, they get downlink control information (DCI) about the preamble of the video data from the eNodeB. The D2D transmitter which is using the same spectrum as LTE then decides on its transmission probability based on the type of the frame mentioned in preamble. If the D2D decides not to transmit when I-packets (from reference frame) are being transmitted, it reduces interference at the eNodeB for LTE uplink transmission. The eNodeB then reports this data-based CQI to the scheduler which then assigns higher MCS to the video transmitting UE until the I-frame packets are being transmitted, This way, it can prevent the I-frame loss. Contrary to this, when a differential frame (B/P) is being transmitted, D2D also transmits and interfere with LTE UE; as a result packet loss happens for differential frames. But since the reference frame is protected, this gives a gain in the received video quality compared to when the strategy is not applied and packet loss happens in reference frame as well.


\section{Simulations \& Numerical Results}
\label{sec:numres}

We assess the performance of the proposed schemes by means of detailed network simulation and processing of exemplar video using NS-3 simulator~\cite{ns3sim} and MATLAB. First, we packetize the video with ffmpeg tool to generate transport stream which is inputted to the NS-3 simulator with a full end-to-end topology over the LTE protocol stack; then based on the interference scenario we generate video packet loss trace which we use in object detection by Computer Vision Toolbox in MATLAB.  We used a H.264/AVC encoded video with high compression to evaluate the performance of video streaming over LTE in presence of D2D. The video is from a surveillance camera with fixed GoP size $128$, where the background is almost fixed and only the foreground changes as objects move. Note that the effectiveness of compression is maximum in this scenario, which is typical of surveillance applications. 

\begin{table}[!t]
\centering
\begin{tabular}{l | l}
\hline
{\bf LTE Parameters} &
{\bf Value} \\ \hline
MAC Scheduler & Proportional Fair (PF)  \\
RLC mode & RLC UM  \\ 
eNodeB power & 25 dB \\
UE Tx power & 23 dB  \\ 
D2D Tx Power & 5 dB \\
Antenna Model & SISO \\
Path Loss model & Friis Propagation Loss \\
Uplink EARFCN & 18100 \\
UE-eNodeB distance & 200 m\\
D2D distance & 10 m \\
\hline
\end{tabular}
\caption{LTE parameters for NS-3 simulation}
\vspace{-1.5em}
\label{tab:tmp}
\end{table}

\subsection{NS-3 Simulation}
We used NS-3 version 3.23 with LTE protocol stack for our simulation. We mimicked D2D communication using the same spectrum and radio channel as of LTE and chose the same bandwidth and central frequency of the channel and same set of sub-channels for both LTE uplink and D2D communication. We consider the UE node is transmitting real-time video streaming data to a remote host via LTE and EPC network. The parameters used for experiments are mentioned in Table~\ref{tab:tmp}. As we consider real-time delay sensitive application, we disable the HARQ and considered RLC UM (Unacknowledge Mode) in NS-3. We will explore more sophisticated content-based HARQ techniques, \emph{e.g}. ~\cite{badia2009}, in future work. In order to map the application packets with the PDUs at MAC layer, we disable the fragmentation and concatenation of the packets in LTE which reduces the overhead and complexity in computation, but does not violate the main idea of interference control. We used channel fading from the NS-3 trace file for Extended Pedestrian A model (EPA) as mentioned in 3GPP standard. Here, we experiment with both low speed (3 kmph) and high speed (10 kmph) fading for which we generate fading trace using MATLAB. For simulating the application, we considered a transport stream of the video with a stream of packets of equal size 188 bytes which we transmit via UDP transport protocol.

\begin{figure}[ht!]
\centering
\begin{minipage}[b]{0.50\textwidth}
	\hspace{-2mm}
	\includegraphics[width=\textwidth]{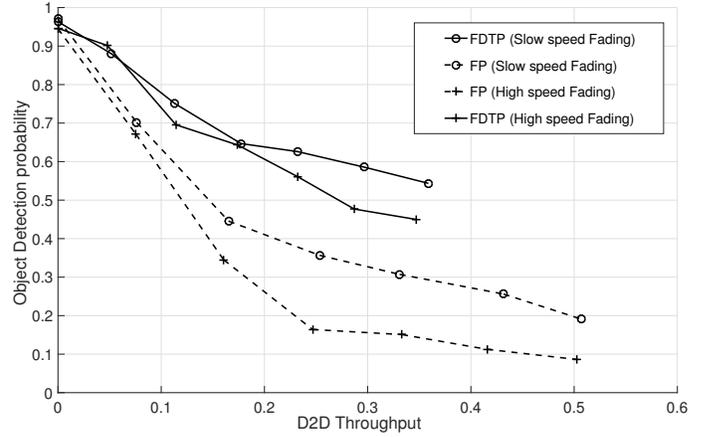}
	\captionsetup{width=1.0 \textwidth}
	\caption{Object detection probability vs D2D throughput in low speed  and high speed fading scenario}
	\label{fig:lowint}
\end{minipage}
 \end{figure}

\begin{figure}[ht!]
\centering
\begin{minipage}[b]{0.50\textwidth}
	\includegraphics[width=\textwidth]{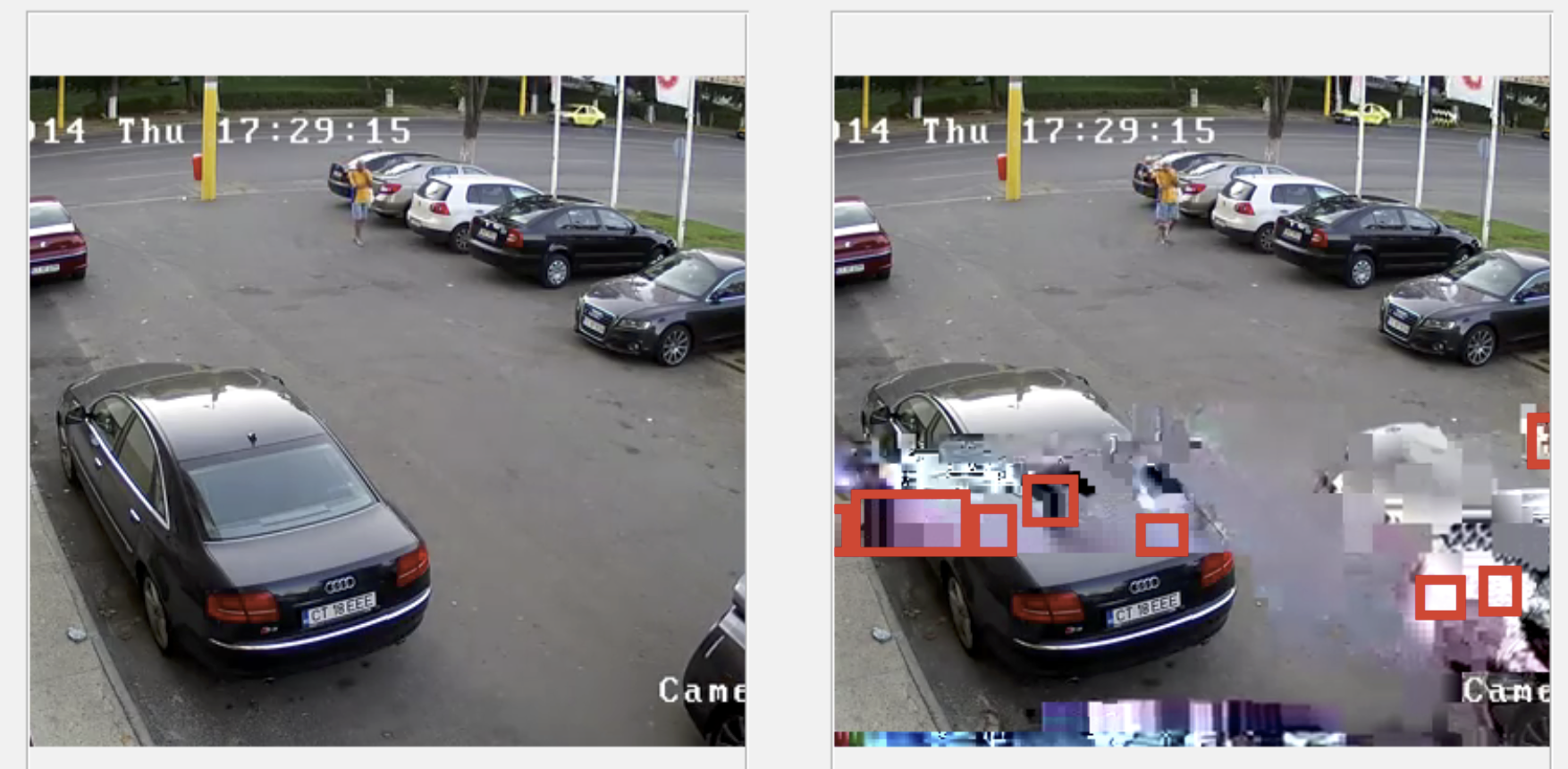}
	\vspace{-1em}
	\caption{Object detection of original and received video}
	\vspace{-1em}
	\label{fig:tfp}
\end{minipage}
\end{figure}

\vspace{-1mm}
\subsection{Results}
We measure the object detection probability of received video with respect to the reference video (in absence of D2D interference) for variation of D2D throughput. Here we considered relative throughput of D2D with respect to reference throughput which assumes D2D always transmits and there is no packet loss. We performed the experiments in high speed fading and low speed fading scenarios. The transmission power of the UE and D2D is fixed throughout the experiment. We remark that these measurements are computed averaging over several random realizations of packet failure sequences.


For a fixed D2D Tx power (5dB), we measure the object detection probability as a function of the D2D throughput in the low speed and high speed fading regimes as shown is Fig.~\ref{fig:lowint}. The lines are obtained by varying the transmission probabilities $\rho$ and $\rho_D$. In FDTP, $\rho_I$ is set to $0$. The results in the Fig.~\ref{fig:lowint} show $20$\% to $30$\%  improvement of the object detection probability in low speed fading case, and almost $25$\% to $35$\%  improvement is obtained in the high-speed fading case. A sample decoded received frame and its original version are shown in the Fig.~\ref{fig:tfp} to visualize the effect of interference on object detection.

\begin{figure}[!t]
	\centering
	\includegraphics[width=\figww]{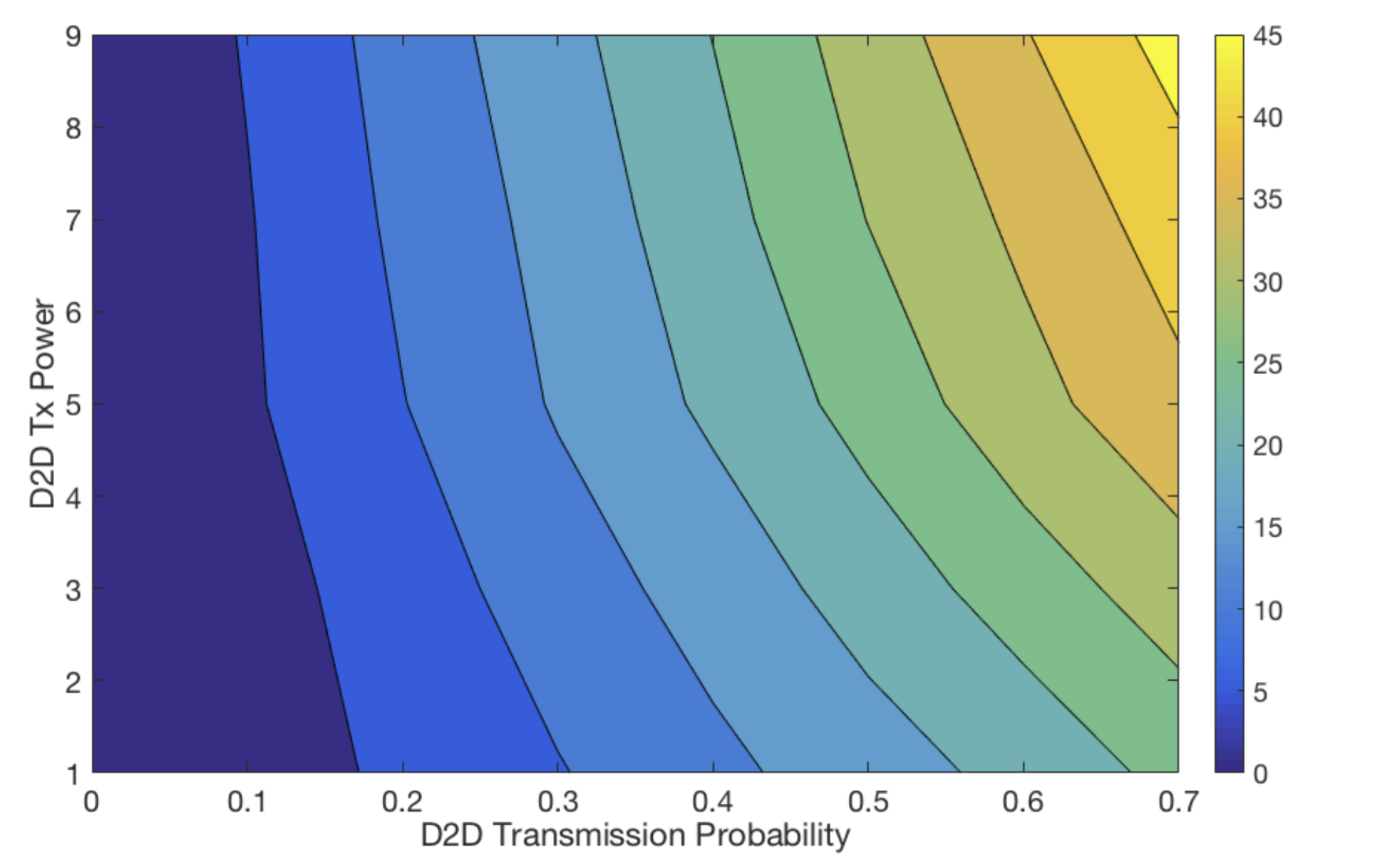}
	\caption{Contour plot for Object detection with respect to D2D transmission probability and D2D transmission power.}
        \label{fig:condet}
\end{figure}

\begin{figure}[!t]
	\centering
	\includegraphics[width=\figww]{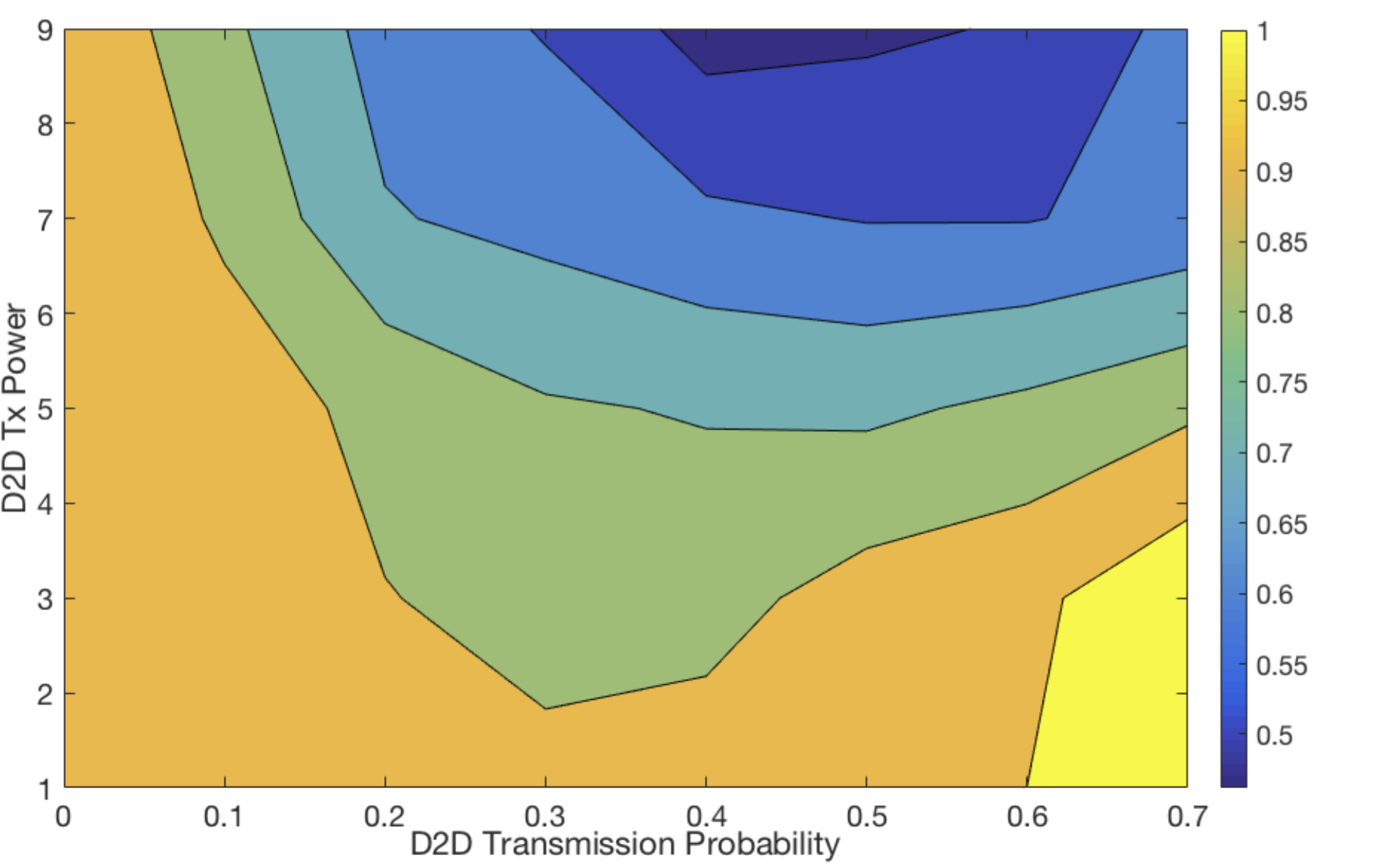}
	\caption{Contour plot for efficiency with respect to D2D transmission probability and D2D transmission power.}
        \label{fig:coneff}
\end{figure}

We also define the efficiency of D2D operations with respect to the degradation caused to the video stream as follows :
\begin{equation}
Efficiency = \frac{\rm Object ~Detection~Probability}{\rm 1 - D2D~Throughput}.
\end{equation}
Fig.~\ref{fig:condet} and ~\ref{fig:coneff} respectively show the throughput of the D2D link and the efficiency as a function of the transmission probability and power. Intuitively, increasing the transmission power or the transmission probability increases the throughput. However, increasing the transmission power or the transmission probability also results in a larger degradation of the object detection probability.
However, Fig.~\ref{fig:coneff} shows that these two parameters can be jointly controlled to minimize the additional impact on the algorithm while increasing throughput. To achieve this, the D2D transmitter should change the parameters in the direction of throughput increase, while remaining in high efficiency regions. Preliminary results indicate that the two parameters can be chosen to generate minimum-impact failure patterns. A full study of this interesting effect, especially in relation with channel statistics is left to future studies.

\section{Conclusions}
\label{sec:ccl}
In this paper we proposed a content-based cognitive transmission strategy for hybrid D2D/LTE networks supporting urban IoT applications. A monitoring application was considered, where video data streams are remotely processed to detect objects. The cognitive strategy shapes the transmission strategy to match the structure of video encoding and processing. Numerical results show significant throughput gain of the D2D link for a given performance degradation of the monitoring application with respect to the case when the interference process is fixed throughout the video transmission.

There are some future scopes to be considered e.g. object detection from a multiple smart camera system and real-time object tracking can be improved by cognitive interference control. Those may include sophisticated computer vision algorithms and intelligent HARQ mechanisms which can further improve real-time applications for public safety and monitoring in urban IoT.
\bibliographystyle{IEEEtran}
\bibliography{IEEEabrv,iot}

\begin{thebibliography}{10}
\providecommand{\url}[1]{#1}
\csname url@samestyle\endcsname
\providecommand{\newblock}{\relax}
\providecommand{\bibinfo}[2]{#2}
\providecommand{\BIBentrySTDinterwordspacing}{\spaceskip=0pt\relax}
\providecommand{\BIBentryALTinterwordstretchfactor}{4}
\providecommand{\BIBentryALTinterwordspacing}{\spaceskip=\fontdimen2\font plus
\BIBentryALTinterwordstretchfactor\fontdimen3\font minus
  \fontdimen4\font\relax}
\providecommand{\BIBforeignlanguage}[2]{{%
\expandafter\ifx\csname l@#1\endcsname\relax
\typeout{** WARNING: IEEEtran.bst: No hyphenation pattern has been}%
\typeout{** loaded for the language `#1'. Using the pattern for}%
\typeout{** the default language instead.}%
\else
\language=\csname l@#1\endcsname
\fi
#2}}
\providecommand{\BIBdecl}{\relax}
\BIBdecl

\bibitem{atzori2010internet}
L.~Atzori, A.~Iera, and G.~Morabito, ``The internet of things: A survey,''
  \emph{Computer networks}, vol.~54, no.~15, pp. 2787--2805, 2010.

\bibitem{IoT}
A.~Zanella, N.~Bui, A.~Castellani, L.~Vangelista, and M.~Zorzi, ``Internet of
  things for smart cities,'' \emph{IEEE Internet of Things Journal}, vol.~1,
  no.~1, pp. 22--32, 2014.

\bibitem{bellavista2013convergence}
P.~Bellavista, G.~Cardone, A.~Corradi, and L.~Foschini, ``{Convergence of MANET
  and WSN in IoT urban scenarios},'' \emph{IEEE Sensors Journal}, vol.~13,
  no.~10, pp. 3558--3567, 2013.

\bibitem{edge_mining_2013}
E.~Gaura, J.~Brusey, M.~Allen, R.~Wilkins, D.~Goldsmith, and R.~Rednic, ``Edge
  mining the internet of things,'' \emph{Sensors Journal, IEEE}, vol.~13,
  no.~10, pp. 3816--3825, Oct 2013.

\bibitem{mitola}
{J.~Mitola}, ``Cognitive radio: an integrated agent architecture for
  software-defined radio,'' Doctor of Technology, Royal Inst. Technol. (KTH),
  Stockholm, Sweden, 2000.

\bibitem{hengstler2007mesheye}
S.~Hengstler, D.~Prashanth, S.~Fong, and H.~Aghajan, ``Mesheye: a
  hybrid-resolution smart camera mote for applications in distributed
  intelligent surveillance,'' in \emph{Proceedings of the 6th international
  conference on Inf. processing in sensor networks}.\hskip 1em plus 0.5em minus
  0.4em\relax ACM, 2007, pp. 360--369.

\bibitem{bramberger2004real}
M.~Bramberger, J.~Brunner, B.~Rinner, and H.~Schwabach, ``Real-time video
  analysis on an embedded smart camera for traffic surveillance,'' in
  \emph{10th IEEE Real-Time and Embedded Technology and Applications
  Symposium}.\hskip 1em plus 0.5em minus 0.4em\relax IEEE, 2004, pp. 174--181.

\bibitem{fitzek2001mpeg}
F.~Fitzek and M.~Reisslein, ``Mpeg-4 and h. 263 video traces for network
  performance evaluation,'' \emph{IEEE Network}, vol.~15, no.~6, pp. 40--54,
  2001.

\bibitem{5350367}
K.~Doppler, M.~Rinne, C.~Wijting, C.~B. Ribeiro, and K.~Hugl,
  ``Device-to-device communication as an underlay to lte-advanced networks,''
  \emph{IEEE Communications Magazine}, vol.~47, no.~12, pp. 42--49, Dec 2009.

\bibitem{3gppps}
``{3GPP} {TS} 23.303, proximity-based services (prose); stage 2 (release 12),
  v.12.0.0,'' February 2014.

\bibitem{yu2009performance}
C.~Yu, O.~Tirkkonen, K.~Doppler, and C.~Ribeiro, ``On the performance of
  device-to-device underlay communication with simple power control,'' in
  \emph{IEEE 69th Vehicular Technology Conference}, 2009, pp. 1--5.

\bibitem{phunchongharn2013resource}
P.~Phunchongharn, E.~Hossain, and D.~Kim, ``Resource allocation for
  device-to-device communications underlaying {LTE}-advanced networks,''
  \emph{IEEE Wireless Communications}, vol.~20, no.~4, pp. 91--100, 2013.

\bibitem{7127550}
Y.~Wen-Bin, M.~Souryal, and D.~Griffith, ``{LTE uplink performance with
  interference from in-band device-to-device (D2D) communications},'' in
  \emph{IEEE Wireless Communications and Networking Conference}, March 2015,
  pp. 669--674.

\bibitem{3gpppr}
``{3GPP} {TR} 36.843 feasibility study on {LTE} device to device proximity
  services - radio aspects,'' 2014.

\bibitem{gunes2014}
T.~T. Gunes, S.~T.~K. U, and H.~Afifi, ``Hybrid model for lte network-assisted
  d2d communications,'' \emph{13th International Conference, ADHOC-NOW 2014,
  Benidorm, Spain, June 22-27, 2014, Pages 100-113}, 2014.

\bibitem{3gpp1}
``{3GPP} {TS} 36.213, {E-UTRA} physical layer procedures.''

\bibitem{chellappa2008}
A.~Sankaranarayanan, A.~Veeraraghavan, and R.~Chellappa, ``Object detection,
  tracking and recognition for multiple smart cameras,'' \emph{Proceedings of
  the IEEE, 96(10), p.1606-1624}, October 2008.

\bibitem{calavia2012}
L.~Calavia, C.~Baladron, J.~Aguiar, B.~Carro, and A.~Esguevillas, ``A semantic
  autonomous video surveillance system for dense camera networks in smart
  cities,'' \emph{Sensors 2012, 12(8), 10407-10429; doi:10.3390/s120810407}.

\bibitem{viola2001}
P.~Viola and M.~Jones, ``Rapid object detection using a boosted cascade of
  simple features,'' \emph{Proceedings of the 2001 IEEE Computer Society
  Conference on Computer Vision and Pattern Recognition, 2001. CVPR 2001.
  (Volume:1 )}, 2001.

\bibitem{dalal2005}
N.~Dalal and B.~Triggs, ``Histograms of oriented gradients for human
  detection,'' \emph{IEEE Computer Society Conference on Computer Vision and
  Pattern Recognition, 2005. CVPR 2005. (Volume:1)}, June 2005.

\bibitem{svm2011}
T.~Malisiewicz, A.~Gupta, and A.~Efros, ``Ensemble of exemplar-svms for object
  detection and beyond,'' \emph{IEEE International Conference on Computer
  Vision (ICCV), 2011}, Nov 2011.

\bibitem{baya2008}
H.~Baya, A.~Essa, T.~Tuytelaars, and L.~V. Gool, ``Speeded-up robust features
  (surf),'' \emph{Computer Vision and Image Understanding, Volume 110, Issue 3,
  June 2008, Pages 346-359}, June 2008.

\bibitem{ns3sim}
``{NS-3} open source network simulator under {GNU} {GPLv2} license,''
  \emph{https://www.nsnam.org/}.

\bibitem{badia2009}
L.~Badia, M.~Levorato, and M.~Zorzi, ``Analysis of selective retransmission
  techniques for differentially encoded data,'' \emph{IEEE International
  Conference on Communications - ICC}, 2009.

\end{thebibliography}

\end{document}